\documentclass[a4paper]{article}
\usepackage{amsmath,amssymb,booktabs}
\usepackage{calc,pstricks,graphicx}

\newcommand{\gw}{gravitational wave}

\newcommand{\ifo}{interferometer}

\newcommand{\msolar}{M_{\odot}}

\usepackage{graphicx}
\usepackage{url}

\begin{document}
\title{Gravitational Wave Astronomy: Delivering on the Promises}
\author{B.F. Schutz\\
{\em School of Physics and Astronomy, Cardiff University} \\{\em Cardiff, U.K.}\\
{\em Max Planck Institute for Gravitational Physics} \\{\em (Albert Einstein Institute)} \\{\em Golm, Germany}
}
\date{}
\maketitle

\begin{abstract}

Now that LIGO and Virgo have begun to detect gravitational wave events with regularity, the field of gravitational wave astronomy is beginning to realise its promise. Binary black holes and, very recently, binary neutron stars have been observed, and we are already learning much from them. The future, with improved sensitivity, more detectors, and detectors like LISA in different frequency bands, has even more promise to open a completely hidden side of the Universe to our exploration.

\end{abstract}

\section{A long road}

The first proposals for kilometer-scale \gw\ \ifo s were formulated in the 1980s, and the scientific justification was based on two principal potential sources: supernova explosions and the inspiral and merger of compact-object binaries, with neutron star and/or black hole components. Since then, the LIGO\cite{LIGOdefpaper}, Virgo\cite{VirgoDef}, and GEO600\cite{GEO600def} detectors have been funded, built, commissioned, and operated from 2005 to 2010 as a network in their initial stage of sensitivity -- without detections at that stage. Also since then, theory and astronomical observations moved forward considerably. Numerical simulations\cite{LRRsupernovasims} showed that the early estimates for gravitational-wave emission from supernovae were overly optimistic. Astronomical observations of binary pulsars and gamma-ray bursts (GRBs) refined the early estimates of binary merger rates, but did not change them much.\cite{LVratespaper}. 

But then came the the detections by the two upgraded Advanced LIGO detectors\cite{aLIGO} of gravitational radiation from coalescing black hole binaries: during the two years before this meeting, four significant events -- named after their detection dates as GW150914\cite{GW150914}, LVT151012, GW151226\cite{GW151226LVT151012}, and GW170104\cite{GW170104} -- have shown that the early estimates for binary sources were, if anything, not optimistic enough. And shortly after the meeting, LIGO and the upgraded Virgo detector\cite{aVirgo} announced not only a further binary black hole detection\cite{GW170814} but, most excitingly, the first detection of a binary neutron star merger\cite{GW170817}, an event that was accompanied by a massive campaign of electromagnetic observations\cite{MMA}. In view of these discoveries, we shall assess  the promise of the very bright new era of gravitational-wave astronomy.

\section{The first detection: GW150914}
The detection of GW150914\cite{GW150914} was a landmark for fundamental physics, astrophysics, and indeed high-precision measurement physics. Its announcement on 11 February 2016 (after a long and careful analysis of the data, and after the discovery papers had been accepted for publication) became a media event, generating press coverage all over the world, tens of millions of Tweets, and such a large number of download requests that the Physical Review Letters website crashed shortly after the announcement. It attracted an unprecedented number of prestige prizes in the first year for the founders of LIGO, Ron Drever, Kip Thorne, and Rai Weiss. And it led to the award of the Nobel Prize for Physics in 2017 \cite{Nobel2017} for Thorne, Weiss, and Barry Barish. (Drever had meanwhile passed away.) 

To understand what this was all about, let us take apart the title and abstract from the discovery paper\cite{GW150914}:
\begin{quotation}
{\bf Observation of Gravitational Waves from a Binary Black Hole Merger}

\ldots

On September 14, 2015 at 09:50:45 UTC the two detectors of the Laser Interferometer Gravitational-Wave Observatory simultaneously observed a transient gravitational-wave signal. The signal sweeps upwards in frequency from 35 to 250 Hz with a peak gravitational-wave strain of $1.0\times10^{-21}$. It matches the waveform predicted by general relativity for the inspiral and merger of a pair of black holes and the ringdown of the resulting single black hole. The signal was observed with a matched-filter signal-to-noise ratio of 24 and a false alarm rate estimated to be less than 1 event per 203 000 years, equivalent to a significance greater than $5.1\sigma$. The source lies at a luminosity distance of $410^{+160}_{-180}$ Mpc, corresponding to a redshift $z = 0.09^{+0.03}_{-0.04}$. In the source frame, the initial black hole masses are $36^{+5}_{-4} \msolar$ and $29^{+4}_{-4} \msolar$, and the final black hole mass is $62^{+4}_{-4} \msolar$, with $3.0^{+0.5}_{-0.5} \msolar c^2$ radiated in gravitational waves. All uncertainties define 90\% credible intervals. These observations demonstrate the existence of binary stellar-mass black hole systems. This is the first direct detection of gravitational waves and the first observation of a binary black hole merger. 

\end{quotation}

The title already conveys the uniqueness of this event: a ``Binary Black Hole Merger''. Not only was the the first merger observed, it was the first {\em direct} evidence of the existence of black holes, the first time we have received any kind of radiation directly generated by the holes themselves. Then there are the holes' remarkable masses: $36^{+5}_{-4} \msolar$ and $29^{+4}_{-4} \msolar$. These are both more massive than {\em any} previously known black holes formed by the evolution of single stars. We know of million- and billion-solar-mass black holes in galaxies and powering quasars, but our knowledge of stellar-mass black holes has come until now exclusively from X-ray astronomy, where we see black holes in binaries paired with normal stars. The standard assumption, therefore, was the binary black holes would be in the same mass-range, although high-mass binaries had been predicted by some theorists.\cite{polishBBH}. GW150914 validated these predictions. 

The event radiated $3.0^{+0.5}_{-0.5} \msolar c^2$  in gravitational wave energy, and when we remind ourselves that most of this energy was emitted in the final 0.2~s of the merger event, then we see that the luminosity was of the order of $3\times10^{55}\,\mathrm{erg}\,\mathrm{s}^{-1}$, or $3\times10^{48}$~W. This is about $10^{22}$ times the luminosity of the Sun! Given that the visible Universe contains roughly $10^{11}$ galaxies, each containing on average roughly $10^{11}$ stars, the luminosity of our single event was competitive with the entire electromagnetic luminosity of the the Universe! Of course, this lasted less than a second, but while it was happening its energy output was astounding. 

This enormous luminosity helps to explain why the signal was detectable despite having travelled $410^{+160}_{-180}$ Mpc (roughly 1.3 billion light-years) to get to us. Even given this enormous distance, the apparent brightness (energy flux) is about $1.5\,\mathrm{mW}\,\mathrm{m}^{-2}$, about 1 lux in the units used for photometry. This is comparable to the brightness of a full Moon! Naturally, all this energy is in gravitational waves, and it couples so weakly to our detectors that we have had to build the most sensitive distance-measuring apparatus ever constructed in order to measure this rather large energy flux. The abstract tells us what sensitivity LIGO needed: ``a peak gravitational-wave strain of $1.0\times10^{-21}$''. The gravitational-wave strain is defined as twice the relative stretching of the LIGO arms as the gravitational wave passed through. This means that, as the wave reached its peak amplitude, LIGO's 4-km arms were oscillating by about 0.1\% of the radius of a proton! 

It is worth noting how remarkable it is that we can know that the distance is 410~Mpc, even with 30\% uncertainty, for an event that has no observed electromagnetic counterpart. This is because such an event is a {\em standard siren}, whose gravitational waveform contains enough information about its intrinsic luminosity to enable us to infer distance from its measured apparent brightness.\cite{Nature86} The relatively large uncertainty in our distance measurement comes from the fact that we observed this event with only two detectors and cannot therefore place it accurately on the sky, nor accurately measure its polarization, both of which we need to compare accurately its apparent brightness with the intrinsic luminosity read from the waveform. Observations with a network of at least three detectors are required for an accurate distance measurement, and that became possible when Advanced Virgo joined the network on 01 August 2017. We will discuss that further in the next section.

The last part of the abstract that I will highlight here is the sentence on significance: ``The signal was observed with a matched-filter signal-to-noise ratio of 24 and a false alarm rate estimated to be less than 1 event per 203 000 years, equivalent to a significance greater than $5.1\sigma$. '' There are two different measures of significance quoted here, and they have quite different statistical meaning. The signal arrived at a time when the detectors were behaving well and the noise was Gaussian. After matched filtering with the best-fitting template from a library of waveforms that incorporate our best understanding of inspiral-and-merger physics,\cite{CBCwaveforms} the combined amplitude signal-to-noise ratio was 24. This is a $24\sigma$ detection, by this measure, and the probability of a false alarm is something like $10^{-90}$, even allowing for the fact that millions of templates would have been tested against the data in the search. But the abstract quotes our confidence at something like $5\sigma$. This much more conservative figure comes from the fact that the detector noise is not always Gaussian, but is characterised by random ``glitches'', which are bursts of interference that originate mainly within the detector. 

The detectors are extremely difficult to control at this sensitivity level; they have a duty cycle only around 60-70\%, and even when in operation mode their Gaussian data is frequently interrupted by these glitches. Since the LIGO detection criterion is that an event should be coincident between the two detectors, the analysis constructs the probability that two randomly occurring glitches could also be coincident, and could look enough like a binary inspiral to masquerade as a signal. On that probability scale, we are confident that a ``glitch event'' as strong as GW150914 would happen by chance less frequently than once per 203 000 years. This is a probability equivalent to that of $5.1\sigma$ in Gaussian noise, but of course the glitch coincidence distribution is not a Gaussian one. 

In fact, this is just a lower limit on the significance against glitches. In all of Advanced LIGO's data during the first observing run (O1) during which this signal was seen, there were no glitches at any time in the two detectors that were strong enough to have masqueraded as this signal even if they had occurred simultaneously, and extrapolating from the glitch distribution suggests that no such glitch coincidence would have occurred even if the detectors had been operated for a time equivalent to the age of the Universe! This was, therefore, an exceptionally strong event to have occurred as the first detection. LIGO would have been confident to have claimed a first detection for a much weaker event than this one.

The report by David Shoemaker in this volume provides a thorough discussion of the LIGO detectors and the first observations.

\section{Further detections}

Given the extraordinary strength of the first detection, it is reassuring that LIGO later found several more that were weaker but still of high significance. Figure~\ref{fig:BHs} shows all the black-hole binary mergers that have been detected to date, and compares their masses with those of known black holes from X-ray binaries. The binaries seen in gravitational waves follow the trend initiated by GW150914, as described above: they are typically much more massive than the black holes known from X-ray observations.  Only GW151226\cite{GW151226LVT151012} has both component masses within the range of X-ray selected black holes.

\begin{figure}[!h]
\centering\includegraphics[width=4in]{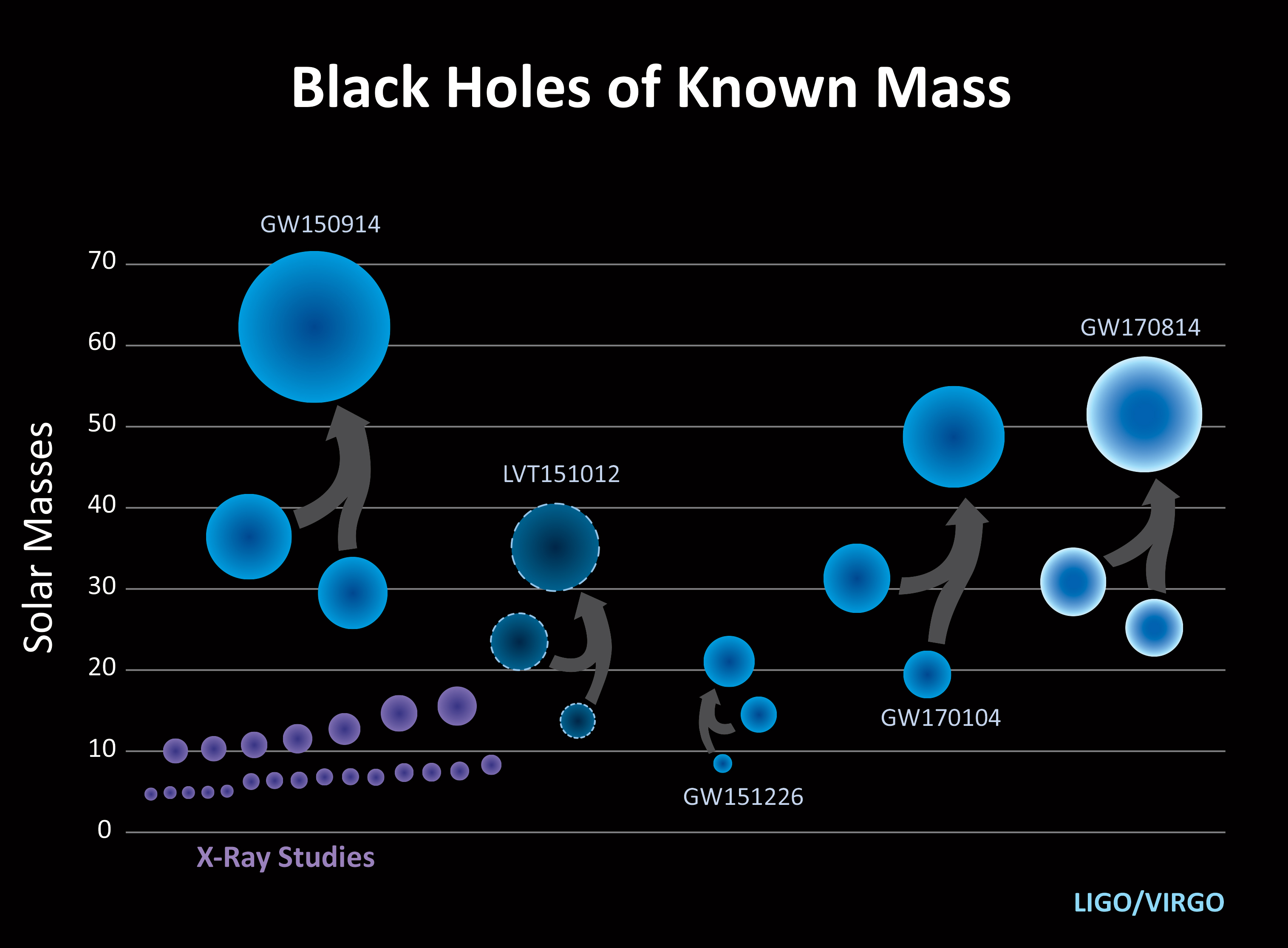}
\caption{A graphic illustrating the masses of all of the binary black hole mergers announced as of the end of October 2017. Both the component masses and the masses of the final remnant are shown. For comparison, the masses of all black holes known from X-ray binary observations are also shown. Credit: National Science Foundation/LIGO/Sonoma State University/A. Simonnet }
\label{fig:BHs}
\end{figure}

The event LVT151012\cite{GW151226LVT151012}, which occurred only four weeks after the GW150914, has the prefix ``LVT'', for LIGO-Virgo-Transient, because its glitch-coincidence significance is not high enough to earn the prefix ``GW'': there is a 5\% chance that it could have been the coincidence of two glitches ($1.7\sigma$). Nevertheless, it seems very likely that this is a real event, since it fits the templates very well, the detectors were otherwise quiet at the time, and its signal-to-noise ratio against the local Gaussian detector noise was 9.7. If real, it was the most distant event so far detected, about 1~Gpc away (redshift 0.2). This distance, combined with its relatively low masses, explains why it was such a weak signal.

The most recent detection shown here, GW170814\cite{GW170814}, was a landmark: it was the first detection made by three interferometers: the two Advanced LIGO detectors, and Advanced Virgo. Virgo had joined the observation run on 1 August 2017, and two weeks later the second-most massive merger ever observed was detected. 

This ensemble of binary black holes has unveiled a fascinating part of the dark Universe. We have been able to look for evidence that the components are spinning black holes; set limits on violations of general relativity; and re-calibrated our understanding of how and how often such black holes are formed. The report by Sathyaprakash in this volume will provide much more detail on them. 

But one further detection needs to be mentioned: GW170817\cite{GW170817}, the binary neutron-star merger event detected just three days after the three-detector black hole event. Again Virgo contributed data, which in this case helped to locate the event sufficiently well on the sky to give electromagnetic observers a good chance to find the expected counterpart. The gravitational-wave data showed that the event was nearby, about 40~Mpc away, which gave added incentive to observers with smaller transient-astronomy survey telescopes. In addition, a weak gamma-ray burst had been observed by the Fermi satellite, just 1.7~s after the end of the gravitational-wave signal, which was the closest known gamma-ray burst. Over 70 different observing partners of LIGO and Virgo responded, locating the counterpart within half a day. The details of all this activity are reported in the first-ever gravitational-wave multimessenger observation paper.\cite{MMA} 

The identification of the host galaxy permitted the accurate standard-siren distance to be combined with optical redshift data to provide a first gravitational-wave measurement of the Hubble constant.\cite{H0} We now have confirmation that some short gamma-ray bursts are produced by neutron-star mergers, and that such mergers have produced much (perhaps most) of the trans-iron heavy elements in the periodic table. Electromagnetic observations are continuing as this report is being written.

The announcement on 16 October 2017 of the detection was another striking media event. 

The LIGO and Virgo detectors were shut down at the end of August 2017 in order to begin a further sensitivity upgrade. The collaboration plans to publish all the results of the observations up to that date in the near future. We can look forward to significantly higher detection rates when the network comes back up in late 2018. The improved sensitivity will lower the Gaussian noise in the detectors and make detections possible over greater volumes. At the same time, having three detectors online will greatly reduce the number of chance glitch events that can masquerade as real ones. So instead of a black-hole event once every month or two, we may be detecting them once every week or so. And although the binary neutron star event GW170817 was probably exceptionally nearby, we can hope for one or more a year in the near future, along with electromagnetic counterpart identifications. The next few years will be rich ones for the science that gravitational-wave observations open up.

\section{Observing with an expanded network after 2020}

Two further projects are underway to build kilometer-scale gravitational-wave detectors: the Japanese KAGRA project\cite{KAGRA} is making steady progress toward an instrument that might take data by 2019, and LIGO-India\cite{LIGOIndia} has received funding and is currently finalizing its choice of site. The addition of these detectors, along with further upgrades in the sensitivity of LIGO and Virgo, could push the event rate up to around 100 per year for black holes, and perhaps nearly that for neutron stars. The loudest event each year could have a signal-to-noise ratio of 60 or more, and we could reach out further than 3~Gpc with our black hole population. The direction-finding of the network will significantly improve, as will the accuracy of the standard-siren distance measurements. In addition, we increase the chance of finally discovering gravitational-wave pulsars.\cite{CW} All of this will lead to exciting new physics and astrophysics.

One of the areas that gravitational-wave detectors are well placed to explore is the nuclear physics of neutron stars. Arguably the most exotic objects in the Universe, neutron stars have densities greater than that of a heavy atomic nucleus, carry magnetic fields whose external strength can be between $10^9$ and $10^{15}$~G (and possibly stronger inside), exhibit superconductivity and superfluidity in their interiors at temperatures in excess of $10^6$~K, have semi-solid crusts, and can rotate faster than $0.1c$ at their surfaces. Observations of binary neutron star mergers can constrain the radii of stars of the measured masses, measure their tidal interactions just before merger, constrain the lifetime of the merger remnant before it finally collapses to a black hole itself -- all of which constrain the nuclear-matter equation of state, inform our understanding of strong-interaction physics, and constrain the properties of the magnetic fields. Observations of continuous waves from spinning, deformed neutron stars can tell us about the physics of the crust and constrain the magnetic field. Observations of short, damped wavetrains from neutron star vibrations can even help us peer into the inner structure of the  star. 

Testing general relativity is a major goal of gravitational wave detection. As we explore the strong-field gravitational physics right up to the horizon, and watch the horizon wobble and shake immediately after a merger, we are in a unique position to look for small deviations from Einstein's theory, whose origin may lie in low-energy effects of quantum gravity. Cosmic censorship is tested by each merger. Joint gravitational and electromagnetic observations can compare the velocity of gravitational waves to that of light. Even distant black-hole mergers constrain the speed of propagation of gravitational waves, because slower-than-light propagation might be accompanied by frequency dispersion, distorting the received signal waveform. A network of 5 detectors can completely constrain the polarization of the waves: if vector, scalar, or longitudinal polarizations are found, this will require a departure from general relativity. 

Astrophysics and cosmology will gain from these observations. The event rate and the mass spectrum of our population of binaries will need to be explained by models of binary formation and evolution. There is every reason to believe that the network could use standard-siren distances to measure the Hubble constant to better than 1\% accuracy, hopefully helping to resolve the tension that now exists between different ways of measuring $H_0$.

Another exciting aspect of having more than three detectors is that the network has more sensitivity to un-modelled sources, for signals that were not predicted, because it can better eliminate the glitch background. I look forward to the discovery of exotic sources like cosmic strings, or even of more prosaic sources that we simply have not imagined.

\section{LISA will open the rich low-frequency band}

Ground-based detectors, even with next-generation 3G sensitivity (described in the next section), will not be able to detect gravitational waves at frequencies below about 1~Hz because the Earth's own gravitational field is noisy enough at those frequencies to mask expected gravitational waves. For this reason, the LISA space-based gravitational wave mission has been under development in Europe and the US since 1995. In December 2015 the European Space Agency launched the LISA Pathfinder (LPF) mission,\cite{LPF} whose aim was to test and prove the special technologies needed for space-based interferometric detection, including drag-free technology. LPF proved to be amazingly successful, exceeding its specification on the drag-free system by factors of more than 10, unequivocally demonstrating that LISA could achieve its design sensitivity or better. The result is that ESA is now fast-tracking LISA for a launch sometime between 2030 and 2034, and NASA has indicated its intention to join as a junior partner. 

With the LPF technology in its three spacecraft, LISA would be able to detect mergers of $10^6\msolar$ black holes (the kinds that seem to inhabit the centers of most ordinary galaxies) with signal-to-noise ratios up to 5000, or at redshifts out to $z=20$. Such mergers are expected in cosmology, as tracers of the mergers of gas clouds that led to our present-day galaxy population. LISA can help elucidate this process. And LISA can also monitor the year-long signals from stellar-mass black holes that happen to spiral into the $10^6\msolar$ holes in relatively nearby galaxies, another process that happens regularly as a result of stellar interactions in the central star clusters of these galaxies. These so-called Extreme Mass-Ratio Inspirals contain a wealth of information about the geometry of the central black hole and are ideal instruments for performing high-precision tests of general relativity. 

The report by Karsten Danzmann in this volume describes LPF and the LISA project in full detail. It is worth noting that other proposals for space-based gravitational wave detectors are being studied in Japan\cite{DECIGO} and China\cite{Taiji,astrod}, so that after LISA there could be further missions.

\section{Ground-based 3.0}

As exciting as the science potential of the current generation of ground-based detectors is, there is a limit to the improvements that can be made in their sensitivity. Proposal are being studied now for what are called third-generation (3G) detectors, which need to be built from new in order to provide sensitivity gains of factors of 10 or more over that of Advanced LIGO. The European Einstein Telescope\cite{ET} study found a workable design with a cryogenically cooled, deep underground triangular 10-km detector array. In the US, above-ground 40-km designs are being studied.\cite{CosmicExplorer} Such detectors will bring our reach for neutron-star binaries out to $z>2$, and we will see every coalescence in the Universe of the types of black holes we have already seen. 

Operating with sophisticated planned transient-astronomy instruments like SKA in the radio, LSST in the optical, and many more, 3G detectors will open up the entire  Universe to gravitational wave astronomy.

\section{Observing in other frequency bands}

The spectrum of interesting gravitational wave frequencies is immense, ranging from periods comparable to the age of the Universe (a primordial stochastic background left over from the end of the inflationary era) up to the 100-Hz and kHz frequencies currently and in future accessible to ground-based instruments. Below LISA's mHz band, radio astronomers are monitoring ultra-stable millisecond pulsars for the tiny changes in pulse arrival times that would be caused by passing gravitational waves. Because these pulsars are only stable enough when averaged over months of observing, this technique of pulsar timing is suited for very low frequencies, 1-10 nHz. These frequencies would be emitted by close binaries of supermassive black holes in the $10^8$ - $10^{10}\msolar$ mass range, which are the engines of quasars. The current observations place interesting limits on a stochastic background due to a large population of binaries of such objects\cite{pulsartiminglimit}, constraining some model predictions.

When the Square-Kilometer Array (SKA) comes online, it will be able to monitor far more pulsars than at present are possible, and it will quickly reach very interesting sensitivities. The detection of individual binary systems of supermassive black holes, and even of some merging systems, will certainly be possible. These observations constrain models of quasar formation and of cosmological structure formation, the latter in much the same way that LISA can, but at later cosmological times when these particular holes have acquired their extremely high masses.

A primordial cosmological background is in principle present at all frequencies, so all gravitational wave detectors can look for it. One difficulty is that it is expected to be very weak, and it might be masked by an astrophysical random background, created by binary systems. However, little is understood about the gravitational waves from inflation, and subsequent stages of early cosmology might well have seen episodes of strong gravitational radiation production. LIGO's unexpectedly high event rate for black holes suggests that their astrophysical background could be particularly strong in the LIGO band, and of course LIGO is searching for a background.\cite{LIGOstoch} Time will tell whether there is a primordial background at LIGO frequencies even larger than that from the binaries. LISA will similarly look for it at mHz frequencies, where a strong background is possible if the electroweak phase transition, which the expanding early Universe is thought to have gone through, produced strong density perturbations. 

But another way to find primordial gravitational waves is to look for them at the beginning: in the cosmic microwave background (CMB). Gravitational waves present at the time that the observed CMB was formed, when photons were decoupling from the charged particles of the primordial plasma, would have left a fingerprint in the polarization of the microwaves we detect today. Several dedicated projects are searching for this fingerprint. Results could come at any time, but it is also possible that the gravitational wave background is much too weak to detect in this way. We simply have to keep looking. 

\section{Gravitational wave and multimessenger \\astronomy}

During the next 20 years we can expect to see a rich science opening up. Big improvements in the number and sensitivity of detectors, including 3G detectors, will bring every black hole binary merger in the Universe into view. LISA will open up the incredibly rich millihertz band, potentially exploring galaxy formation out to redshifts of 20 or more. With SKA, pulsar timing will come into its own as a way of observing supermassive black hole binaries. And we may well be able to detect a primordial gravitational wave background, either in the CMB or at higher frequencies with one or more of our instruments. 

Now that gravitational wave detection has begun to deliver on the promises made over the last decades, we can  feel optimistic about its promise in the future. It is important to see ground-based and space-based detectors as part of a wider network of multimessenger transient-astronomy observational instruments, covering the electromagnetic spectrum and including neutrino detectors. Gravitational wave astronomy will become an integral and indispensable part of observational astronomy, and will further increase the importance and contribution of astronomy and cosmology to our understanding of fundamental physics.

\vskip6pt








\end{document}